\documentclass[12pt]{article}

\usepackage{graphicx}
\graphicspath{{./figs/}} 

\usepackage[letterpaper,margin=1in]{geometry}

\renewenvironment{abstract}
	{\quotation}
	{\endquotation}

\date{}

\makeatletter
\renewcommand{\fnum@figure}{\textbf{Figure \thefigure}}
\renewcommand{\fnum@table}{\textbf{Table \thetable}}
\makeatother

\usepackage{url}

\usepackage{subcaption}

\usepackage{amsmath,dsfont,amsfonts,braket}
\usepackage{amssymb} 

\usepackage{hyphenat} 

\usepackage{microtype} 

\usepackage[dvipsnames]{xcolor}         
\usepackage[unicode=false,psdextra]{hyperref}
\hypersetup{
    pdfdisplaydoctitle,
    bookmarksnumbered=true,
    bookmarksopen,
    breaklinks,
    linktoc=all,
    plainpages=false,
    unicode=true,
    colorlinks=true,                  
    allcolors=black,     
    linkcolor=NavyBlue,      
    citecolor=NavyBlue
}

\usepackage[backend=biber,
    autocite=superscript,
    url=false,
    sorting=none,
    hyperref=true,
    isbn=false,
    citestyle=numeric-comp]{biblatex} 
\usepackage[noabbrev,nameinlink,capitalise]{cleveref} 

\crefformat{equation}{(#2#1#3)}
\crefrangeformat{equation}{(#3#1#4) to~(#5#2#6)}
\crefmultiformat{equation}{(#2#1#3)}%
{ and~(#2#1#3)}{, (#2#1#3)}{ and~(#2#1#3)}

\crefname{SI}{Supplementary Section}{Supplementary Section} 

\crefname{SubFig_a}{Figure}{Figures}
\crefname{SubFig_b}{Figure}{Figures}
\crefname{SubFig_c}{Figure}{Figures}
\crefname{SubFig_d}{Figure}{Figures}
\crefname{SubFig_e}{Figure}{Figures}
\crefname{SubFig_f}{Figure}{Figures}
\crefname{SubFig_bc}{Figures}{Figures}
\crefname{SubFig_cd}{Figures}{Figures}
\crefname{SubFig_ce}{Figures}{Figures}
\crefname{SubFig_ab}{Figures}{Figures}
\crefname{SubFig_ac}{Figures}{Figures}
\crefname{SubFig_df}{Figures}{Figures}

\creflabelformat{SubFig_a}{#2#1(a)#3}
\creflabelformat{SubFig_b}{#2#1(b)#3}
\creflabelformat{SubFig_c}{#2#1(c)#3}
\creflabelformat{SubFig_d}{#2#1(d)#3}
\creflabelformat{SubFig_e}{#2#1(e)#3}
\creflabelformat{SubFig_f}{#2#1(f)#3}
\creflabelformat{SubFig_bc}{#2#1(b-c)#3}
\creflabelformat{SubFig_cd}{#2#1(c-d)#3}
\creflabelformat{SubFig_ab}{#2#1(a-b)#3}
\creflabelformat{SubFig_ce}{#2#1(c-e)#3}
\creflabelformat{SubFig_ac}{#2#1(a-c)#3}
\creflabelformat{SubFig_df}{#2#1(d-f)#3}

\newcommand{\MyFigLabel}[1]{
  \label{#1}
  \label[SubFig_a]{#1_a}
  \label[SubFig_b]{#1_b}
  \label[SubFig_c]{#1_c}
  \label[SubFig_d]{#1_d}
  \label[SubFig_e]{#1_e}
  \label[SubFig_f]{#1_f}
  }

\newcommand{\MyFigLabelRange}[3]{\label[SubFig_#2#3]{#1_#2#3}}

\usepackage[export]{adjustbox}

\newcommand{\mytilde}{\raise.17ex\hbox{$\scriptstyle\mathtt{\sim}$}}

\def\mytitle{
	Crystal-phase quantum dots in AlGaAs nanowires
}
\title{\bfseries \boldmath \mytitle}

\usepackage{authblk} 
\author[1]{Rohan~Radhakrishnan}
\author[2]{Rodion~Reznik}
\author[3]{Gilles~Patriarche}
\author[1]{Lorenzo~Leandro}
\author[2]{Igor~Ilkiv}
\author[2]{Anna~Andreeva}
\author[4]{Artem~Khrebtov}
\author[2,4]{George~Cirlin}
\author[1]{Nika~Akopian*}
\affil[1]{\small DTU Department of Electrical and Photonics Engineering, Technical University of Denmark, 2800 Kongens Lyngby, Denmark.}
\affil[2]{\small Faculty of Physics, St. Petersburg State University, Universitetskaya Embankment 7-9, 199034 St. Petersburg, Russia.}
\affil[3]{\small Centre de Nanosciences et Nanotechnologies, Université Paris Saclay, CNRS, 91120 Palaiseau, France.}
\affil[4]{\small Department of Epitaxial nanotechnologies, Alferov University, Khlopina 8/3, 194021 St. Petersburg, Russia.}

\date{\small *Corresponding author. Email: nikaak@dtu.dk}

\makeatletter
\let\MyAuthorsList\@author
\let\MyDate\@date
\makeatother

\begin{document}

\newrefsection

\maketitle

\begin{abstract} \bfseries \boldmath
    Crystal-phase quantum dots (CPQDs)—quantum dots in nanowires defined by crystal structure rather than material composition—con\-sti\-tute the only platform capable of fabricating quantum-dot arrays with the ultimate precision of a single atomic layer. This intrinsic control yields perfectly aligned quantum dots with atomically sharp interfaces, providing a unique pathway toward scalable quantum-dot-based photonic quantum technologies. To date, CPQDs have been studied primarily in binary semiconductors, such as InP and GaAs, where their emission linewidths are typically in the meV range, thereby limiting their technological potential. Here, we report, for the first time, CPQDs in AlGaAs nanowires and show bright single-photon emission with linewidths as narrow as 61~\textmu eV and low background emission, demonstrating optical quality well beyond typical CPQDs. We attribute this performance to a type-I band alignment, as suggested by an exciton lifetime of 1 ns, significantly shorter than that typically observed in type-II CPQDs. Additionally, we observe an exciton fine-structure splitting and a Zeeman splitting, as commonly observed in standard type-I self-assembled quantum dots.
\end{abstract}


\vspace{10mm}

\noindent Owing to their unique optical properties, quantum dots (QDs) have emerged as essential building blocks for photonic quantum technologies \autocite{heindelQuantumDotsPhotonic2023,senellartHighperformanceSemiconductorQuantumdot2017}. They became excellent sources of indistinguishable single photons and entangled photon pairs, and offer an efficient interface between photons and the spin of a trapped charge, all of which are essential resources for quantum photonic computing, quantum communication, and quantum networks. The applications, however, are limited to single- or two-quantum-dot devices \autocite{jenningsSelfAssembledInAsGaAs2020} because the most efficient QDs are grown using methods such as Stranski-Krastanov growth, which are inherently stochastic and offer limited control over the design \autocite{scheibnerOpticallyMappingElectronic2008}.

On the other hand, crystal-phase quantum dots (CPQDs) feature atomically sharp interfaces and can be vertically stacked with perfect alignment along the nanowire axis. They are defined by crystal structure—switch between zincblende and wurtzite phases—rather than material composition. The band offset between the two phases induces a spatial variation in the band structure along the nanowire axis, thereby enabling the formation of QDs without altering the material composition \autocite{akopianCrystalPhaseQuantum2010}. The perfect alignment and ultimate precision of a single atomic layer position CPQDs as the only platform capable of fabricating complex multi-quantum-dot arrays with atomic-layer precision, thereby opening the way to a new class of scalable quantum-dot-based photonic quantum technologies, such as multi-qubit photonic devices \autocite{taherkhaniHighfidelityOpticalQuantum2019,hastrupAllopticalChargingCharge2020,liLocationQubitsMultipleQuantumDot2024a,hacklExperimentalProposalProbe2023}.

Despite the unique advantages of CPQDs, two major challenges must still be addressed to unlock their technological potential. The first is to transition from randomly formed CPQDs to deterministic fabrication with control over position and size, and significant progress has recently been made in elucidating the growth mechanism and achieving precise atomic-scale control of the crystal phase along the nanowire \autocite{jacobssonInterfaceDynamicsCrystal2016b,pancieraPhaseSelectionSelfcatalyzed2020,geijselaersAtomicallySharpCrystal2021,vainoriusConfinementThicknessControlledGaAs2015}. The second challenge concerns the optical quality of their emission, which typically lags behind that of conventional compositionally defined QDs. Mostly studied in binary semiconductors like GaAs and InP, reported CPQDs exhibit low brightness, linewidths on the order of a few millielectronvolts, and substantial spectral overlap with background emission from the rest of the nanowire \autocite{akopianCrystalPhaseQuantum2010,geijselaersAtomicallySharpCrystal2021,loitschCrystalPhaseQuantum2015,spirkoskaStructuralOpticalProperties2009}. Only one example of bright and clean sub-millielectronvolts single-photon emission from InP CPQDs can be found \autocite{bouwesbavinckPhotonCascadeSingle2016b}, showing its feasibility but also the difficulty in obtaining high-quality emission.

Here, we investigate, for the first time, CPQDs formed in AlGaAs nanowires and demonstrate their single-photon statistic as well as the remarkably high optical quality of their emission compared to typical CPQDs. We observe clean and spectrally sharp exciton and biexciton emission lines with linewidths of 104~\textmu eV and 61~\textmu eV, respectively. 

We then examine the origin of this outstanding optical quality compared to other CPQDs.

Two major constraints that typically degrade CPQD performance are the presence of unintended other CPQDs in the nanowires, which contribute to background emission and charge noise by acting as charge traps when excited above bandgap \autocite{dalacuUltracleanEmissionInAsP2012a}, and the intrinsically lower emission rate and higher charge noise sensitivity associated with the type-II band alignment. The first one depends on the growth conditions, and our nanowires exhibit a relatively low density of phase switching (15 to 50~zincblende insertions/\textmu m). The second one arises from the electron-hole spatial separation in type-II band alignment, which reduces the overlap of their wavefunctions and consequently lowers the radiative recombination rate. Moreover, type-II QDs are more sensitive to Stark shift and, therefore, to charge noise, due to the appearance of a permanent dipole moment, as a result of the electron-hole spatial separation with asymmetric repartition of the wavefunction \autocite{janssensStarkShiftSingle2002, ramanathanQuantumconfinedStarkEffects2013}. We could also expect the permanent dipole moment to increase the efficiency of CPQDs as a source of charge noise.

Unlike most CPQDs, our AlGaAs CPQDs show a type-I band alignment, as suggested by their 1~ns exciton lifetime, which is much shorter than the tens-of-nanosecond lifetimes typically measured in type-II CPQDs \autocite{akopianCrystalPhaseQuantum2010, spirkoskaStructuralOpticalProperties2009, bouwesbavinckPhotonCascadeSingle2016b}. Additionally, polarization-resolved and magneto-photoluminescence measurements reveal a fine-structure splitting and a Zeeman splitting, as commonly observed in standard type-I self-assembled QDs.

By demonstrating the excellent optical quality of CPQDs in AlGaAs nanowires, this work paves the way toward CPQDs that meet the performance requirements of quantum technologies, while offering a unique pathway—thanks to their atomically sharp demarcation and perfect vertical alignment—to scalable quantum-dot-based devices, where state-of-the-art QDs are limited.

\section*{Results}

    \subsection*{Core-shell emission of AlGaAs nanowires}

\begin{figure}[htbp]
\includegraphics[width=0.9\textwidth, center]{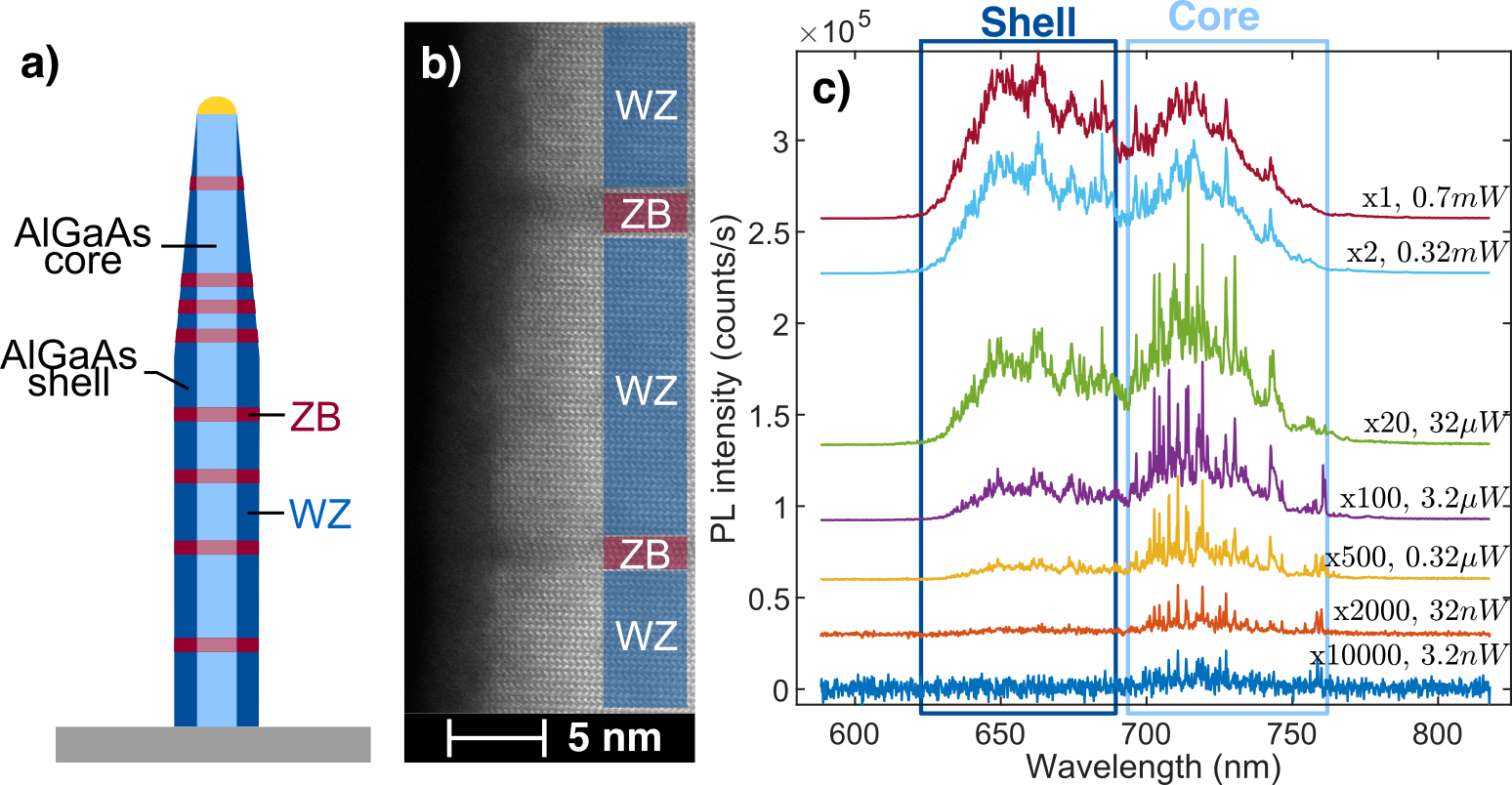}
\caption{\textbf{Core-shell structure and emission.} \textbf{(a)} Schematic of the AlGaAs nanowire structure. \textbf{(b)} HAADF-STEM image of two zincblende (ZB) insertions. \textbf{(c)} Macro-photoluminescence as a function of the above-band-gap excitation power. Multiple nanowires are probed simultaneously. The emission spectral range can be separated into two regions, depending on whether the emission comes from the core or the shell.
}
\MyFigLabel{fig:AlGaAs_CPQD_Fig1}
\end{figure}

We start by conducting a structural analysis of the wurtzite AlGaAs nanowires and the zincblende insertions that form the QDs using high-angle annular dark-field scanning transmission electron microscopy (HAADF-STEM), and by identifying the distinct emission spectral ranges (\Cref{fig:AlGaAs_CPQD_Fig1}). The nanowires are Au-catalyzed and grown by molecular beam epitaxy (MBE) with a nominal AlAs/GaAs ratio of 0.4 \autocite{cirlinAlGaAsAlGaAsGaAs2017a}. They spontaneously develop into a core-shell structure (\Cref{fig:AlGaAs_CPQD_Fig1_a}), where the aluminium content in the core is lower than in the shell, thereby reducing the bandgap in the core relative to the shell. The catalyst droplet size determines the core diameter, which ranges from 7 to 20~nm. Short zincblende insertions appear randomly along the nanowire with a density that varies from 15 to 50~insertions/\textmu m depending on the sample.

The atomic-resolution STEM image of \Cref{fig:AlGaAs_CPQD_Fig1_b} displays the typical size of the zincblende insertions, which is on the order of a few atomic layers (< 3~nm). The crystal phase of the core propagates to the shell; therefore, a zincblende insertion in the core is also present in the shell.

Quantum emitters in the core and the shell emit on distinct spectral ranges due to the difference in aluminium content \autocite{leandroWurtziteAlGaAsNanowires2020,barettinDirectBandGap2023a} (\Cref{fig:AlGaAs_CPQD_Fig1_c}). At low power, only a few nanowires are excited, and we observe multiple sharp emission lines originating from the CPQDs formed by zincblende insertions in the core. At higher power, the emitters in the core saturate, and emission from the shell at higher energy becomes predominant. Our study focuses here on CPQDs formed in the core, emitting from 700 to 750~nm; emission from the shell will be the focus of future studies.

    \subsection*{Core crystal-phase quantum dots free of compositional fluctuation}

\begin{figure}[htbp]
\includegraphics[width=0.85\textwidth, center]{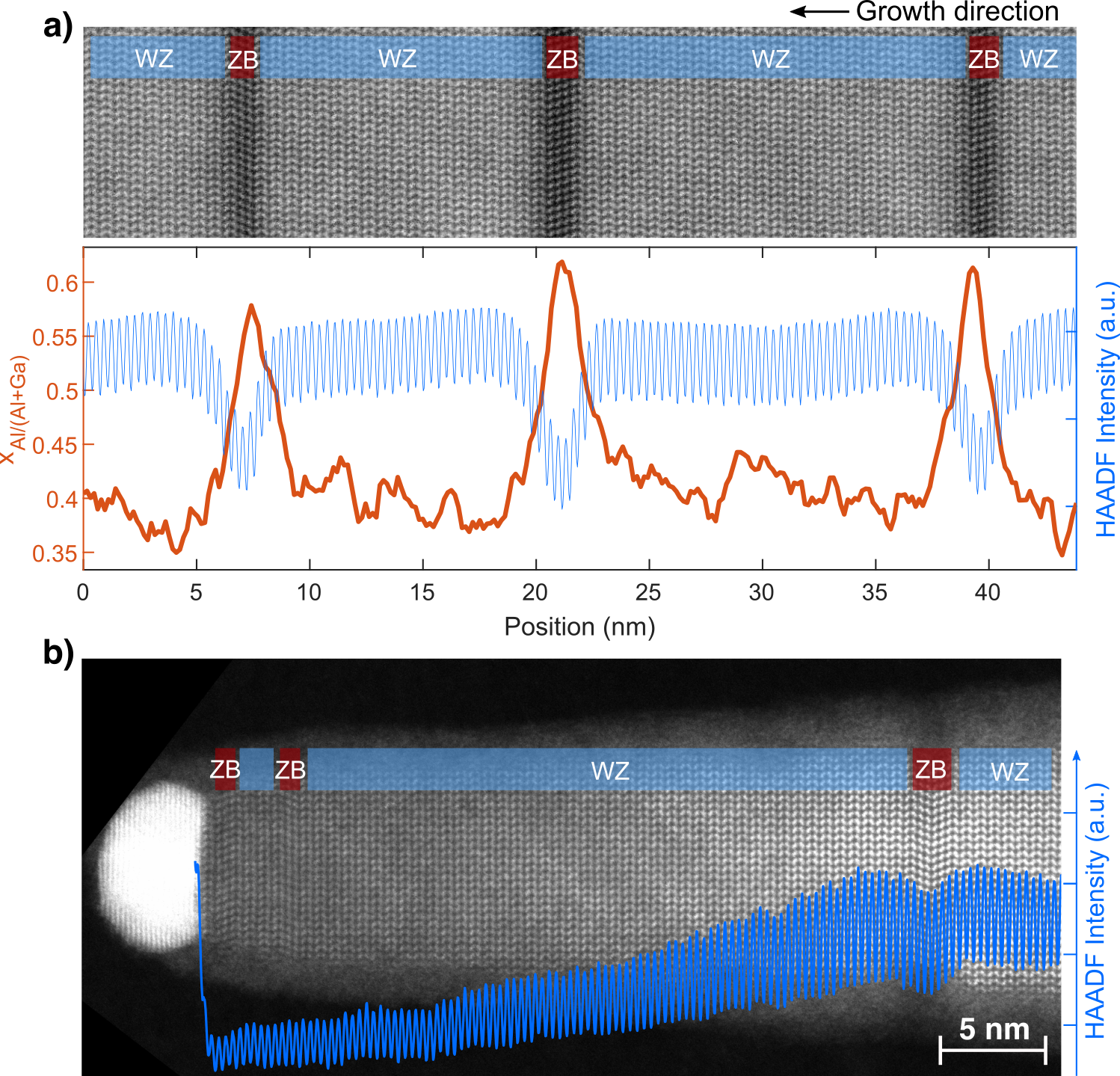}
\caption{\textbf{Compositional analysis of the crystal-phase insertions.} \textbf{(a)} HAADF-STEM of the middle of the nanowire, where the signal originates predominantly from the thick shell (\mytilde 70~nm thickness). The brightness drops at the locations of the three zincblende (ZB) insertions due to a local increase in aluminium content. The extracted HAADF intensity, integrated over the image's pixel columns, is plotted in blue together with the x\textsubscript{Al/(Ga+Al)} ratio (in red) derived from an EDX linescan acquired on the same area. \textbf{(b)} HAADF-STEM of the top of the nanowire. The HAADF intensity (in blue), sensitive to compositional fluctuations, shows no variation in aluminium content at the two least shell-covered zincblende insertions near the catalyst droplet.
}
\MyFigLabel{fig:AlGaAs_CPQD_EDX}
\end{figure}

We then analyze the atomic composition of the zincblende insertions, which, unlike binary compounds, can vary with the growth thermodynamics of the two phases. Energy-dispersive X-ray spectroscopy (EDX) measured on the Al\textsubscript{x}Ga\textsubscript{1-x}As shell shows consequent increases from x = 0.4 in the wurtzite phase up to x = 0.6 in the zincblende insertions (\Cref{fig:AlGaAs_CPQD_EDX_a}). To accurately measure Al/Ga fluctuations in the core, we use the variation in contrast of HAADF-STEM at the tip of the nanowires \autocite{prianteSharpeningInterfacesAxial2016}, where the shell is the thinnest (\Cref{fig:AlGaAs_CPQD_EDX_b}). HAADF-STEM is more sensitive than EDX, with an atomic sensitivity of 0.1~\% \autocite{pantzasExperimentalQuantificationAtomicallyresolved2021a}, and the HAADF intensity decreases with an increase in aluminium content, as shown in \Cref{fig:AlGaAs_CPQD_EDX_a} when compared to EDX. In \Cref{fig:AlGaAs_CPQD_EDX_b}, the HAADF intensity increases continuously with increasing distance from the catalyst droplet, as the nanowire thickens. The profile shows no intensity drop at the locations of the two zincblende insertions near the catalyst droplet, where the shell is almost nonexistent, indicating no compositional fluctuation in the core due to a phase change. For comparison, we note a clear drop in intensity at the location of the third zincblende insertion, farther from the catalyst droplet, which we attribute to an increase in the shell's aluminium content. EDX shows a variation from $\mathrm{x = 0.38 \pm 0.03}$ on the zincblende insertion to $\mathrm{x = 0.34 \pm 0.01}$ on the wurtzite segment located on its right (see \Cref{Energy-dispersive X-ray spectroscopy map}).

Therefore, we conclude that aluminium fluctuations due to crystal-phase switches appear only in the shell and not in the core. This difference between the core and the shell is explained by the two growth modes (lateral and axial), which have distinct dynamics. Thus, CPQDs in the AlGaAs core are not "hidden" compositional QDs, and their confinement potential is determined by the band offset between zincblende and wurtzite, as CPQDs in binary compounds.

    \subsection*{Narrow-linewidth single-photon emission and type-I band alignment}

We now demonstrate the single-photon nature and high-quality emission of the CPQDs in the AlGaAs core by focusing on an individual emitter.

\Cref{fig:AlGaAs_CPQD_Fig2} shows one example exhibiting bright exciton and biexciton photoluminescence emission with narrow linewidths of respectively 104 and 61~\textmu eV ($\mathrm{\pm 4}$~\textmu eV), with low background emission. Its excitation power dependence shows the typical exciton-biexciton behavior, with the exciton line appearing at low power, followed by the biexciton line, which becomes predominant as the exciton emission saturates and then decreases.

The origin of each line is further confirmed by the substantial bunching observed on the cross-correlation measurement (\Cref{fig:AlGaAs_CPQD_Fig2_c})—a clear sign of a cascade emission pathway between the biexciton and exciton state. Because of the limited time resolution of our setup ($\mathrm{FWHM \gtrsim  505~ps}$), we do not resolve the expected antibunching dip for negative delays associated with detecting an exciton photon followed by a biexciton photon (see \Cref{Impact of the time resolution on the correlation measurements} for a discussion on the impact of the time resolution).

Nevertheless, the autocorrelation measurement of the exciton line reveals a clear antibunching dip with $\mathrm{g^{(2)}(0) < 0.5}$, demonstrating the single-photon nature of the emission (\Cref{fig:AlGaAs_CPQD_Fig2_d}). The dip depth is again limited by the resolution of the setup, and \Cref{Other examples of AlGaAs crystal-phase quantum dots} provides an example of high single-photon purity ($\mathrm{g^{(2)}(0) < 0.1}$) measured with a better time resolution ($\mathrm{FWHM \approx 110~ps}$).

From the fitting of the coincidence counts of \Cref{fig:AlGaAs_CPQD_Fig2_d}, we measure an antibunching time of $\mathrm{\tau_a = 0.90~ns}$, which agrees well with the fast component of the lifetime ($\mathrm{\tau_1 = 1~ns}$) measured by time-resolved photoluminescence (\Cref{fig:AlGaAs_CPQD_Fig2_b}). We attribute the slow exponential decays to excitons recaptured in the QD, originating from a long-standing reservoir. Identifying its nature would require further study.

The short lifetime and antibunching time observed here are in agreement with other measurements on AlGaAs CPQDs (\Cref{Other examples of AlGaAs crystal-phase quantum dots} presents two other examples). Such timescales are shorter than the typical lifetimes, ranging from 3 to 10~ns, for InP and GaAs type-II CPQDs \autocite{akopianCrystalPhaseQuantum2010, spirkoskaStructuralOpticalProperties2009, bouwesbavinckPhotonCascadeSingle2016b}, where spatial separation of the electron and hole wavefunction produces few-nanosecond lifetimes. A shorter dynamics strongly suggests a type-I band alignment, where electrons and holes reside in the same segment, as observed in type-I structures like GaAs  \autocite{leandroResonantExcitationNanowire2020} and InGaAs QDs \autocite{radhakrishnanAlGaAsNanowiresUniversal2026} embedded in similar AlGaAs nanowires, which exhibit lifetimes of 1~ns or shorter. 

Continuing on the same AlGaAs CPQD, polarization-resolved and magneto\hyp{}photoluminescence (\Cref{fig:AlGaAs_CPQD_Fig3}) show additional signs typical of a type-I confinement. The alternating polarization orientations reveal a fine-structure splitting of the exciton state (\Cref{fig:AlGaAs_CPQD_Fig3_a}), which is commonly observed in QDs and originates from the electron-hole exchange interaction and anisotropic confinement \autocite{bayerFineStructureNeutral2002}. The large value of 235~\textmu eV measured here again suggests a type-I band alignment, because a lower value would be expected for type-II QDs, due to the electron-hole spatial separation\autocite{klenovskyExcitonicFineStructure2022}. Furthermore, under a magnetic field, Zeeman interaction splits the exciton state, accompanied by a diamagnetic shift (\Cref{fig:AlGaAs_CPQD_Fig3_b}). The spectral lines split into two, as is usually observed in standard type-I QDs \autocite{bayerFineStructureNeutral2002}, and this result significantly differs from what has been previously measured on type-II CPQDs, where the emission lines split into four distinct lines \autocite{akopianCrystalPhaseQuantum2010}.

\begin{figure}[htbp]
\includegraphics[width=0.9\textwidth, center]{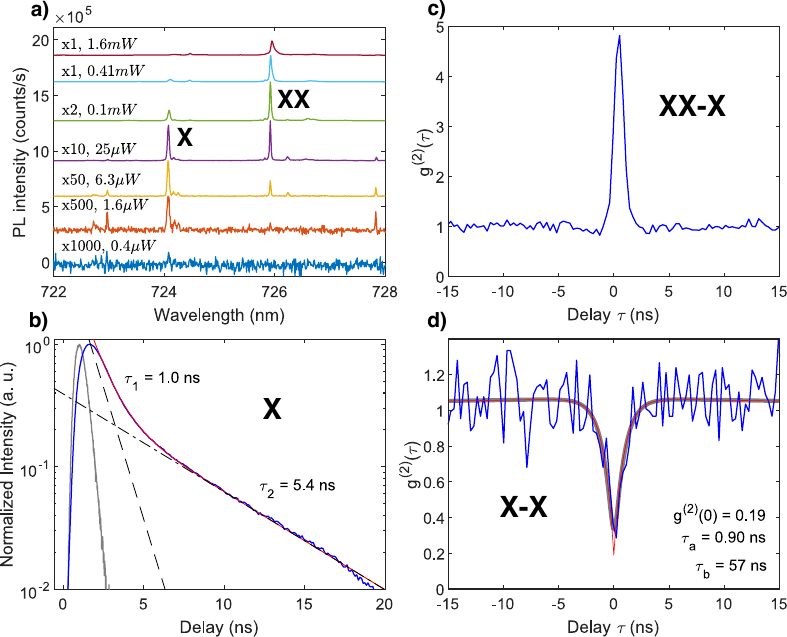}
\caption{\textbf{Exciton - biexciton complex in an AlGaAs CPQD.} \textbf{(a)} Power-dependent micro-photoluminescence showing exciton (X) and biexciton (XX) states. Only emission with vertical polarization is acquired (see \Cref{fig:AlGaAs_CPQD_Fig3} for horizontal polarization). \textbf{(b)} Time-resolved photoluminescence of the exciton line. The decay is fitted (red) with the sum of two decaying exponentials, which are also plotted separately in dashed lines. The extracted characteristic times are plotted next to each decay. \textbf{(c)} Cross-correlation between the exciton and biexciton emission lines showing a strong bunching consistent with cascade emission. For positive delay, the start channel is the biexciton emission and the stop channel is the exciton emission; the opposite order holds for negative delay. \textbf{(d)} Autocorrelation measurement of the exciton line showing antibunching. The data are fitted with a two-exponential $g^{(2)}(\tau)$ function accounting for antibunching (characteristic time $\tau_a$) and power bunching (characteristic time $\tau_b$). The transparent grey line is the fit to the data, and the red line corresponds to the same fit partially corrected for the non-ideal instrument response function (they are very close to each other). See \Cref{Fitting functions for the autocorrelation measurements,Impact of the time resolution on the correlation measurements} for details of the fit.
}
\MyFigLabel{fig:AlGaAs_CPQD_Fig2}
\MyFigLabelRange{fig:AlGaAs_CPQD_Fig2}{c}{d}
\end{figure}

\begin{figure}[htbp]
\includegraphics[width=0.9\textwidth, center]{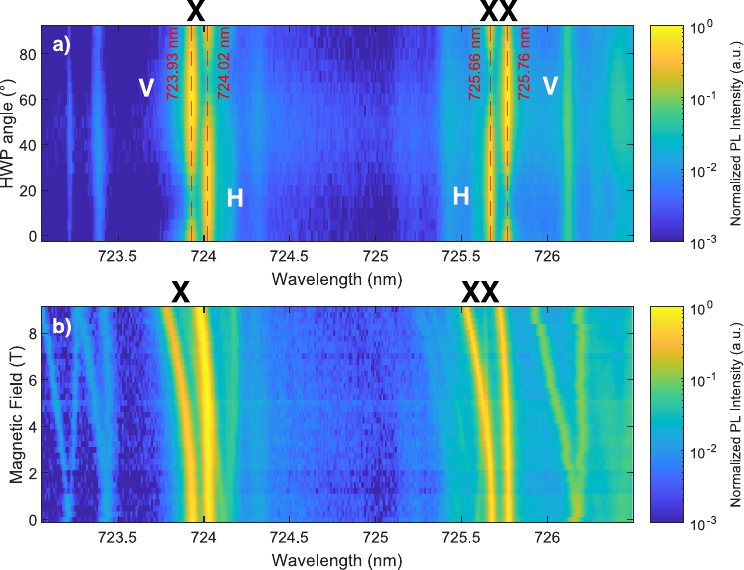}
\caption{\textbf{Fine-structure of the exciton state and its magnetic evolution.} \textbf{(a)} Linear-polarization-resolved photoluminescence showing a clear switching between horizontal (H) and vertical (V) polarization of the exciton and biexciton lines, revealing an exciton fine-structure splitting of $\mathrm{235 \pm 30 ~\mu eV}$. \textbf{(b)} Magneto-photoluminescence showing Zeeman effect and diamagnetic shift. The magnetic field is parallel to the nanowire's axis (Faraday configuration).
}
\MyFigLabel{fig:AlGaAs_CPQD_Fig3}
\end{figure}

\section*{Discussion and conclusion}

In summary, we have demonstrated that short zincblende insertions into the core of a wurtzite AlGaAs nanowire create CPQDs with remarkable optical quality, including high single-photon purity, narrow exciton and biexciton linewidths, and low background emission. We attribute this superior optical performance relative to typical CPQDs to a type-I band alignment and to the low density of insertions. 

To our knowledge, only one other bright, spectrally narrow single-photon emission from CPQDs with low background has been previously reported. In this study, Bouwes Bavinck et al. \autocite{bouwesbavinckPhotonCascadeSingle2016b} measured clean 23~\textmu eV-linewidth exciton emission from InP CPQDs, where the band alignment is type-II as confirmed by the \mytilde10~ns lifetime. A low density of zincblende insertions is also reported (15~insertions/\textmu m), which is on the lower end of our AlGaAs nanowires' range (15 to 50~insertions/\textmu m). The higher purity could explain the spectrally sharp emission, owing to a lower charge-noise environment with fewer charge traps \autocite{dalacuUltracleanEmissionInAsP2012a}, whereas our AlGaAs CPQDs are more robust to charge noise thanks to the type-I band alignment. 

Additionally, our findings provide the first insight into the band alignment between the zincblende and wurtzite phases in AlGaAs, as it has not yet been predicted or measured, despite the extensive use of AlGaAs as a protective shell or host material and the crucial role of band alignment for modeling and designing novel structures and devices. For this matter, the effect of the nanowire's core diameter should be considered, as reducing the diameter increases radial quantum confinement and has been predicted to lead to a transition from type-II to type-I band alignment in InP CPQDs \autocite{zhangWideInPNanowires2010} and has also been observed \autocite{loitschCrystalPhaseQuantum2015} and modeled \autocite{climenteElectronsHolesExcitons2016} in CPQDs in thin GaAs nanowires (6-13 nm core diameter \autocite{loitschCrystalPhaseQuantum2015}). Our nanowires have similar diameters (7-20 nm), and such a transition is likely to happen here as well.

Finally, the demonstration of two-photon cascade emission and Zeeman splitting, together with the bright, clean, and spectrally narrow single-photon emission, shows the potential of CPQDs in AlGaAs nanowire as performant QDs for quantum technologies. Two-photon cascade emission is an essential step toward generating polarization-entangled \autocite{akopianEntangledPhotonPairs2006,versteeghObservationStronglyEntangled2014,pennacchiettiOscillatingPhotonicBell2024} or time-bin-entangled \autocite{jayakumarTimebinEntangledPhotons2014} photon pairs, and Zeeman splitting enables fine-tuning of the QD's energy levels, a crucial advantage for interfacing it with other devices in a hybrid quantum network \autocite{akopianArtificialAtomLocked2013,leandroNanowireQuantumDots2018,laneveQuantumTeleportationDissimilar2025} or creating spin qubits \autocite{warburtonSingleSpinsSelfassembled2013}. Most importantly, thanks to the atomically sharp demarcation and perfect alignment, high-quality emission from CPQDs paves the way for fabricating QD arrays with atomic precision—a unique platform for a new class of scalable photonic quantum technologies \autocite{taherkhaniHighfidelityOpticalQuantum2019,liLocationQubitsMultipleQuantumDot2024a,hastrupAllopticalChargingCharge2020,hacklExperimentalProposalProbe2023}.

\section*{Methods}

\paragraph{Nanowire growth}

The AlGaAs nanowires were grown by MBE using the Au-catalyzed vapor-liquid-solid (VLS) method on Si(111) substrates. Before the growth, we remove the native oxide from the substrate surface by wet chemical treatment in an HF:H\textsubscript{2}O solution, followed by annealing at 850°C in the metallization chamber, where we then deposit a \mytilde0.5~nm-thick Au layer at 550°C. The substrate was then transferred into the growth chamber while maintaining the vacuum. After reaching the substrate growth temperature (510°C) and stabilization of As\textsubscript{4} flux onto the surface, the Al and Ga sources are simultaneously opened to initiate the AlGaAs nanowires growth under As-rich conditions for 25~minutes with a nominal AlAs/GaAs ratio of 0.4.

\paragraph{HAADF-STEM and EDX}

The HAADF-STEM images were acquired on an aberration-corrected FEI TITAN 200 TEM/STEM microscope operating at 200~keV, with a probe convergence half-angle of 17.6~mrad and detection inner and outer half-angles of 69~mrad and 200~mrad, respectively.

The nanowires were transferred onto a carbon membrane supported by a copper grid and oriented along the <11-20> zone axis of the wurtzite structure, corresponding to the <110> zone axis of the cubic zinc blende structure, thereby enabling unambiguous discrimination between the structures.

All micrographs were acquired for 41~s at 2048 by 2048 pixels with an 8~\textmu s dwell. EDX measurements were performed using, in addition, a Chemistem system with a collection angle of 0.8 sr and a Bruker windowless Super-X four-quadrant detector. The linescan was acquired for 10~min, and EDX chemical maps were acquired over 20-40~min, with drift correction performed using cross-correlation. The EDX profiles are quantified using the Cliff-Lorimer method, with absorption effects corrected for light elements such as aluminium. Reference standards were used to refine the k-factors for quantification \autocite{pantzasExperimentalQuantificationAtomicallyresolved2021}.

\paragraph{Photoluminescence, polarization-resolved measurements and magnetic field}

Photoluminescence measurements were carried out at low temperatures (1.5-6 K) using a 532 nm continuous-wave green laser as the excitation source. The emitted light was analyzed with a 750 mm focal-length spectrometer equipped with a low-noise, Peltier-cooled CCD detector. The spectrometer entrance slit was set to 30 µm, and the CCD pixel size was 20 µm.

Macro-photoluminescence was collected using a 5x objective with a numerical aperture (NA) of 0.12, whereas micro-photoluminescence measurements on individual nanowires employed a 100x 0.65NA objective. High-resolution, polarization-resolved, and magneto-photoluminescence measurements utilized an 1800 gr/mm holographic grating, yielding a spectral resolution of approximately 0.02 nm.

For linear-polarization-resolved measurements, a half-wave plate placed before a polarizing beam splitter was rotated to select the desired polarization. A second half-wave plate was positioned in front of the spectrometer to compensate for the grating's polarization sensitivity and to maximize detection efficiency.

Magneto-photoluminescence experiments were performed with the sample mounted in a closed-cycle cryostat equipped with a 9~T vertical superconducting magnet.

\paragraph{Photon correlation and time-resolved photoluminescence measurements}

Second-order correlation measurements were performed under continuous-wave excitation (532~nm laser), and two different setups with different time resolutions were used to measure emission. For \Cref{fig:AlGaAs_CPQD_Fig2,fig:AlGaAs_CPQD_QD3}, the emission was first sent to a 50:50 beam splitter (Hanbury Brown and Twiss setup) and then to two monochrometers (spectrometer mentioned above with 1800~gr/mm) to filter out the spectral lines. The light was then coupled to multimode fibers and sent to avalanche photodiode (APD) single-photon detectors, connected to a time tagger. The timing jitter (FWHM) of the detectors is 350~ps and 72~ps for a single channel of the time tagger, which gives an overall jitter of 505~ps. This jitter has not been measured directly but is estimated based on the specifications and test reports provided by the suppliers. The actual overall timing jitter is likely higher (see \Cref{Impact of the time resolution on the correlation measurements}).

Time-resolved photoluminescence measurements were performed using a 440~nm pulsed laser diode operating at 20~MHz repetition rate and with a 200~ps FWHM pulse width.


\section*{Acknowledgments}

N. Akopian acknowledges financial support for the research of this work from the European Research Council (grant No. 101003378) and the Carlsberg Foundation (grant No. CF22-1543). G. Patriarche acknowledges financial support for the C2N characterization platform from the French technology network Renatech. R. Reznik and G. Cirlin acknowledge support of the Russian Science Foundation (grant No. 25-79-10101) for the growth of the samples.

\section*{Author contributions}

R. Radhakrishnan and N. Akopian conceived the study. R. Reznik, I. Ilkiv, A. Andreeva, A. Khrebtov, and G. Cirlin grew samples. G. Patriarche provided the HAADF-STEM data and structural analysis. R. Radhakrishnan and L. Leandro performed the optical experiments and analyzed the data. All authors discussed the results. R. Radhakrishnan wrote the manuscript with inputs from all co-authors. N. Akopian coordinated the project.


\printbibliography


\newpage


\renewcommand{\thefigure}{S\arabic{figure}}
\renewcommand{\thetable}{S\arabic{table}}
\renewcommand{\theequation}{S\arabic{equation}}
\renewcommand{\thepage}{S\arabic{page}}
\setcounter{figure}{0}
\setcounter{table}{0}
\setcounter{equation}{0}
\setcounter{section}{0}
\setcounter{page}{1} 

\begin{center}
\newrefsection 

\section*{Supplementary Materials for\\ \mytitle}

{\large
\MyAuthorsList
}\\
\MyDate
\end{center}

\subsubsection*{This PDF file includes:}
Supplementary Text\\
Figures S1 to S3\\

{\hypersetup{linkcolor=black}
\tableofcontents}

\newpage

\section{Energy-dispersive X-ray spectroscopy map}
\label[SI]{Energy-dispersive X-ray spectroscopy map}

\Cref{fig:AlGaAs_CPQD_EDXMap} presents the results from the energy-dispersive X-ray spectroscopy (EDX) map of the nanowire tip shown in the main text (\Cref{fig:AlGaAs_CPQD_EDX_b}). The aluminium content rises as one moves farther away from the catalyst droplet, due to increased shell thickness. At the zincblende insertion in region of interest (ROI) number 4, we note a greater increase in aluminium content than in ROI 5, due to the zincblende phase of the shell. ROI 6, where the second-closest zincblende insertion to the catalyst droplet is located, does not show higher aluminium content. Those results are in perfect agreement with the HAADF intensity profile presented in the main text.

\begin{figure}[htbp]
\includegraphics[width=\textwidth, center]{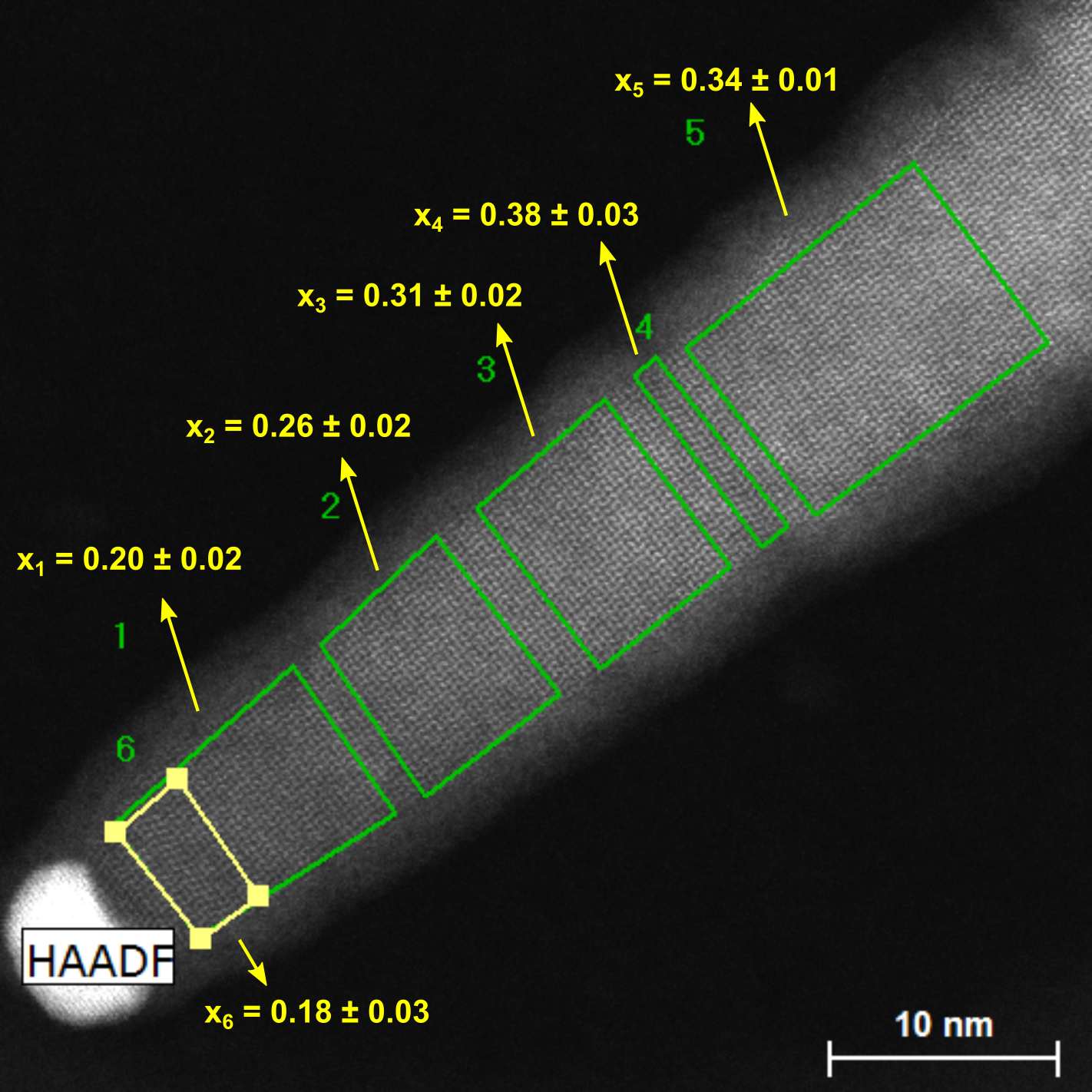}
\caption{\textbf{EDX map of the nanowire tip presented in the main text.} Dark-field scanning transmission electron microscopy (HAADF-STEM) image with superimposition of the different regions of interest (ROI) labeled from 1 to 6 (green rectangles). On each ROI, we measure the average content of aluminium, gallium, and arsenic, from which we calculate the AlAs/GaAs ratio x indicated in yellow next to each ROI.
}
\MyFigLabel{fig:AlGaAs_CPQD_EDXMap}
\end{figure}

\section{Other examples of AlGaAs crystal-phase quantum dots}
\label[SI]{Other examples of AlGaAs crystal-phase quantum dots}

\begin{figure}[htbp]
\includegraphics[width=\textwidth, center]{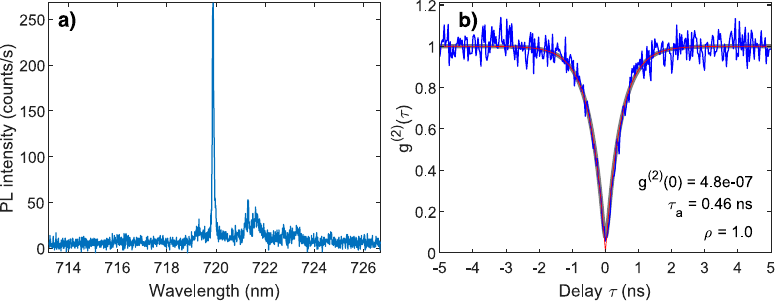}
\caption{\textbf{Second example of AlGaAs CPQD emission.} \textbf{(a)} High-resolution micro-photoluminescence showing a sharp emission line ($\mathrm{FWHM = 180 \pm 4}$~\textmu eV) from a CPQD in the core. \textbf{(b)} Autocorrelation measurement showing the single-photon statistic of the emission. The data are fitted with a $g^{(2)}(\tau)$ function accounting for the antibunching (characteristic time $\tau_a$). The transparent grey line is the fit to the data, and the red line is the same fit corrected for the non-ideal instrument response function (they are very close). See \Cref{Fitting functions for the autocorrelation measurements} for details of the fit. The second-order correlation setup used here differs from that in \Cref{fig:AlGaAs_CPQD_Fig2_cd} of the main text and in \Cref{fig:AlGaAs_CPQD_QD3}. It offers better time resolution, with an overall jitter (FWHM) of 110~ps thanks to the use of fast superconducting nanowire single-photon detectors (SNSPDs). The spectral line was filtered out of the emission spectrum thanks to a tunable reflective Bragg filter. The light was then coupled into a single-mode fiber, split in two by a 50:50 fiber-based beam splitter, and finally sent to SNSPDs connected to the time tagger.
}
\MyFigLabel{fig:AlGaAs_CPQD_QD2}
\end{figure}

\begin{figure}[htbp]
\includegraphics[width=\textwidth, center]{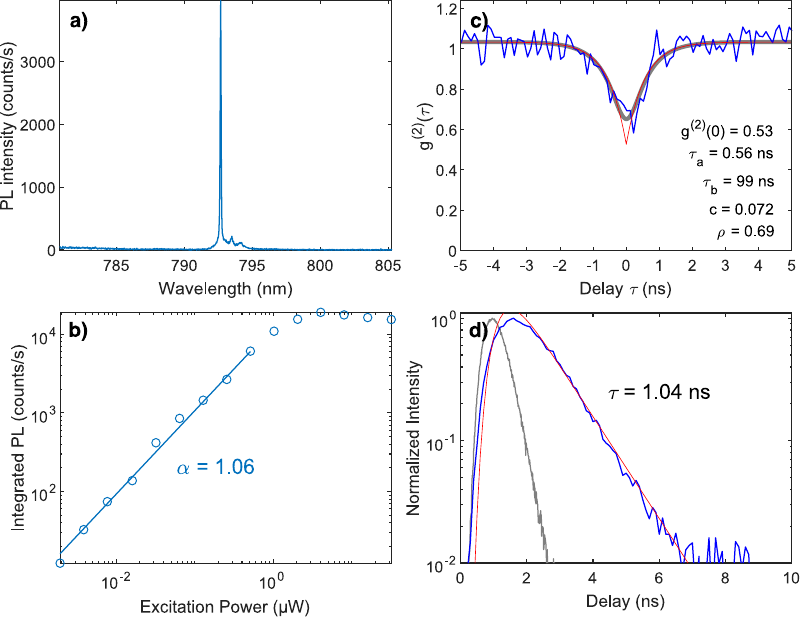}
\caption{\textbf{Third example of AlGaAs CPQD emission.} \textbf{(a)} High-resolution micro-photoluminescence showing a sharp emission line ($\mathrm{FWHM = 168 \pm 4}$~\textmu eV) from a CPQD in the core. \textbf{(b)} Integrated photoluminescence over the FWHM of the emission line as a function of the excitation power. The intensity $I$ below saturation is fitted with the power law $I \propto Power^{\alpha}$. \textbf{(c)} Autocorrelation measurement showing a small antibunching dip, which is limited by the time resolution of the setup (see \Cref{Impact of the time resolution on the correlation measurements}). The data are fitted with a two-exponential $g^{(2)}(\tau)$ function accounting for antibunching (characteristic time $\tau_a$) and power bunching (characteristic time $\tau_b$). The transparent grey line is the fit to the data, and the red line is the same fit partially corrected for the non-ideal instrument response function. See \Cref{Fitting functions for the autocorrelation measurements,Impact of the time resolution on the correlation measurements} for details of the fit. \textbf{(d)} Time-resolved photoluminescence (blue) and mono-exponential fit of the decay. The fit accounts for the instrument response function measured with the pulsed laser (grey).
}
\MyFigLabel{fig:AlGaAs_CPQD_QD3}
\end{figure}

\Cref{fig:AlGaAs_CPQD_QD2,fig:AlGaAs_CPQD_QD3} \footnote{The nanowire measured in \Cref{fig:AlGaAs_CPQD_QD3} is from a different sample than the one used for the optical study (main text and \Cref{fig:AlGaAs_CPQD_QD2}). It also features an embedded InGaAs segment within the AlGaAs nanowire; however, we focus on the AlGaAs emission, which is well-separated from the InGaAs emission (810-900~nm). The spectral range of the AlGaAs core emission is redshifted in this sample because of a slightly lower aluminium content.} present two other examples of crystal-phase quantum dot (CPQD) emission in the core of AlGaAs nanowire. In both cases, the emission exhibits a sharp spectral line with a linewidth of \mytilde 175~\textmu eV and single-photon statistics. Moreover, the fit returns a short antibunching time $\tau_a$ (0.46~ns and 0.56~ns). In addition, the time-resolved photoluminescence of \Cref{fig:AlGaAs_CPQD_QD3_d} shows a short lifetime of 1~ns that we can associate with the recombination of an exciton or charge exciton because of the linear dependency of the photoluminescence with excitation power \autocite{finleyChargedNeutralExciton2001} (\Cref{fig:AlGaAs_CPQD_QD3_b}).

\section{Fitting functions for the autocorrelation measurements}
\label[SI]{Fitting functions for the autocorrelation measurements}

The raw data obtained from the autocorrelation measurements are coincidence counts, which we normalized by the average rate obtained at a long delay time ($\mathrm{\tau \approx 500~ns}$), where the light can be considered uncorrelated. For \Cref{fig:AlGaAs_CPQD_QD2}, we then fit the normalized data with the following $g^{(2)}$ function convoluted with the instrument response function (see next section) \autocite{regelmanSemiconductorQuantumDot2001,brouriPhotonAntibunchingFluorescence2000}:

$$ g^{(2)}(\tau) = 1 - \rho^2 \cdot e^{-\frac{|\tau - \tau_0|}{\tau_a}} $$

where $\tau_a$ is the antibunching time, $\rho$ is the signal-to-noise ratio, and $\tau_0$ is a fixed delay between the two paths of the Hanbury Brown and Twiss interferometer. The x-axis has been shifted to set the origin to $\tau_0$, making this value not visible on the figure.

\bigskip

For the autocorrelation of \Cref{fig:AlGaAs_CPQD_Fig2_d} of the main text and \Cref{fig:AlGaAs_CPQD_QD3_c}, we add a bunching term \autocite{regelmanSemiconductorQuantumDot2001}:

$$ g^{(2)}(\tau) = 1 + \rho^2 \left( c \cdot e^{-\frac{|\tau - \tau_0|}{\tau_b}} - (1 + c) \cdot e^{-\frac{|\tau - \tau_0|}{\tau_a}} \right) $$

where $\tau_b$ is the bunching time and $c$ is the bunching amplitude.

\section{Impact of the time resolution on the correlation measurements}
\label[SI]{Impact of the time resolution on the correlation measurements}

The antibunching dips measured in \Cref{fig:AlGaAs_CPQD_Fig2_d} of the main text and in \Cref{fig:AlGaAs_CPQD_QD3_c} do not reach zero, due to the low time resolution of the setup (jitter $\mathrm{FWHM \gtrsim  505~ps}$). This limitation is confirmed by the autocorrelation presented in \Cref{fig:AlGaAs_CPQD_QD2_b} where, despite a similar antibunching time, the value for $g^{(2)}(0)$ is close to zero, thanks to the better time resolution of the setup employed there (jitter $\mathrm{FWHM \approx 110~ps}$), which uses faster single-photon detectors.

To compensate for this limitation, we take into account the instrument response function (IRF) in the fit, which we approximate by a Gaussian of FWHM equal to the jitter. The final fitting function is the convolution of the above $g^{(2)}$ functions with the IRF. We note that despite the convolution with the IRF, the corrected $g^{(2)}$ function (red line) still does not reach 0 at zero delay and stays very similar to the uncorrected function (transparent grey line). The reason is that we likely underestimate the overall jitter (i.e., overestimate the time resolution) because the jitter is calculated from the supplier's specifications. A higher jitter value would reduce the $g^{(2)}(0)$ value, the signal-to-noise ratio $\rho$ and the antibunching time $\tau_a$ returned from the fit.

For the cross-correlation in \Cref{fig:AlGaAs_CPQD_Fig2_c} of the main text, we expect an antibunching dip at negative delays; however, the dip is unresolved, due to the limited time resolution.

\clearpage 
\printbibliography

\end{document}